\documentclass[
reprint,
 amsmath,amssymb,
 aps,
prd,
]{revtex4-2}
\usepackage{hyperref}
\usepackage{lineno}
\usepackage{float}
\usepackage{makecell}
\hypersetup{
hypertex=true,
colorlinks=true,
linkcolor=blue,
anchorcolor=blue,
citecolor=blue,
urlcolor=blue
}
\usepackage{rotating}
\usepackage{placeins} 
\usepackage{graphicx}
\usepackage{dcolumn}
\usepackage{bm}
\usepackage{amsmath}  
\usepackage{lmodern}  
\newcommand{\BESIIIorcid}[1]{\href{https://orcid.org/#1}{\hspace*{0.1em}\raisebox{-0.45ex}{\includegraphics[width=1em]{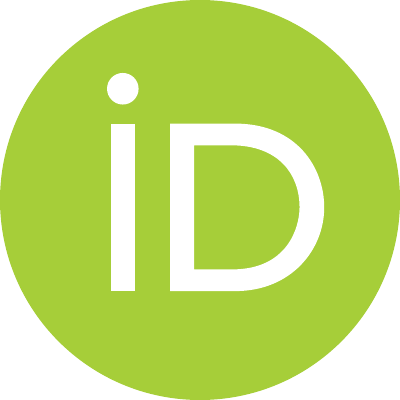}}}} 

\begin{document}

\preprint{APS/123-QED}

\title{Search for the isospin-violating decays $\bm{\chi_{cJ}\to\Lambda\bar{\Sigma}^{0}+c.c.}$ and $\bm{\eta_{c}\to\Lambda\bar{\Sigma}^{0}+c.c.}$}

\author{
\begin{center}
M.~Ablikim$^{1}$\BESIIIorcid{0000-0002-3935-619X},
M.~N.~Achasov$^{4,c}$\BESIIIorcid{0000-0002-9400-8622},
P.~Adlarson$^{81}$\BESIIIorcid{0000-0001-6280-3851},
X.~C.~Ai$^{86}$\BESIIIorcid{0000-0003-3856-2415},
C.~S.~Akondi$^{31A,31B}$\BESIIIorcid{0000-0001-6303-5217},
R.~Aliberti$^{39}$\BESIIIorcid{0000-0003-3500-4012},
A.~Amoroso$^{80A,80C}$\BESIIIorcid{0000-0002-3095-8610},
Q.~An$^{77,64,\dagger}$,
Y.~H.~An$^{86}$\BESIIIorcid{0009-0008-3419-0849},
Y.~Bai$^{62}$\BESIIIorcid{0000-0001-6593-5665}
% ...
\\
\vspace{0.2cm}
(BESIII Collaboration)\\
\vspace{0.2cm} {\it
$^{1}$ Institute of High Energy Physics, Beijing 100049, People's Republic of China\\
% ...
$^{4}$ Budker Institute of Nuclear Physics SB RAS (BINP), Novosibirsk 630090, Russia\\
% ...
$^{31}$ INFN Sezione di Ferrara, (A)INFN Sezione di Ferrara, I-44122, Ferrara, Italy; (B)University of Ferrara, I-44122, Ferrara, Italy\\
% ...
$^{39}$ Johannes Gutenberg University of Mainz, Johann-Joachim-Becher-Weg 45, D-55099 Mainz, Germany\\
% ...
$^{62}$ Southeast University, Nanjing 211100, People's Republic of China\\
% ...
$^{64}$ State Key Laboratory of Particle Detection and Electronics, Beijing 100049, Hefei 230026, People's Republic of China\\
% ...
$^{77}$ University of Science and Technology of China, Hefei 230026, People's Republic of China\\
% ...
$^{80}$ University of Turin and INFN, (A)University of Turin, I-10125, Turin, Italy; (B)University of Eastern Piedmont, I-15121, Alessandria, Italy; (C)INFN, I-10125, Turin, Italy\\
$^{81}$ Uppsala University, Box 516, SE-75120 Uppsala, Sweden\\
% ...
$^{86}$ Zhengzhou University, Zhengzhou 450001, People's Republic of China\\

\vspace{0.2cm}
$^{\dagger}$ Deceased\\
% ...
$^{c}$ Also at the Novosibirsk State University, Novosibirsk, 630090, Russia\\
% ...

}

\end{center}
}

\begin{abstract}
Using a sample of $(2712.4\pm14.3)\times10^{6}$~$\psi(3686)$ events collected with the BESIII detector,  we perform a search for the isospin-violating decays $\chi_{cJ}\to\Lambda\bar{\Sigma}^{0}+c.c.~(J=0,1,2)$ and $\eta_{c}\to\Lambda\bar{\Sigma}^{0}+c.c.$
No significant signal for $\chi_{cJ}$ or $\eta_{c}$ is observed in the $\Lambda\bar{\Sigma}^{0}$ invariant mass distribution.  The upper limits on the branching fractions at the 90\% confidence level are set  to be
$\mathcal{B}$($\chi_{c0}\to\Lambda\bar{\Sigma}^{0}+c.c.$)$<1.5\times 10^{-6}$,
$\mathcal{B}$($\chi_{c1}\to\Lambda\bar{\Sigma}^{0}+c.c.$)$<1.6\times 10^{-6}$,
$\mathcal{B}$($\chi_{c2}\to\Lambda\bar{\Sigma}^{0}+c.c.$)$<1.7\times 10^{-6}$ and
$\mathcal{B}$($\eta_{c}\to\Lambda\bar{\Sigma}^{0}+c.c.$)$<6.2\times 10^{-5}$ for the first time. 
\end{abstract}
\maketitle

\section{INTRODUCTION}
Theoretically, SU(3) flavor symmetry plays an important role in calculating the branching fractions of charmonium decaying into baryon–antibaryon pairs~\cite{BaldiniFerroli:2019abd,Liu:2010um,Kivel:2022fzk,Dai:2023vsw}.
A particularly interesting case is the  $\psi(2S)\to\Lambda\bar{\Sigma}^{0}$ decay, which was initially reported with a large branching fraction~\cite{Ferroli:2020xnv,Dobbs:2017hyd}, suggesting either isospin violation or data reliability issues~\cite{Rosini:2025wfm}.  However, subsequent precise measurements by BESIII~\cite{BESIII:2021mus} resolved these discrepancies, conclusively excluding the unexpected significant isospin violation effects~\cite{Rosini:2025wfm}. 

The color octet mechanism (COM)~\cite{PhysRevD.51.1125}  proposed nearly three decades ago for P-wave heavy quarkonia decays, has been extensively applied to predict exclusive  $\chi_{cJ}$ ($J=0,1,2$) decays to baryon-antibaryon pairs~\cite{Ping:2004sh,Liu:2010um}.
However, the discrepancies between theoretical predictions and the experimental results on the branching fractions of $\chi_{c0}\to\Lambda\bar{\Lambda}$ and $\chi_{c0}\to\Sigma^{0}\bar{\Sigma}^{0}$~\cite{BESIII:2012ghz} 
indicate the  violation of the established helicity selection rule~\cite{Brodsky:1981kj}.

Recent theoretical work ~\cite{PhysRevD.110.056007} has predicted  $\chi_{cJ}\to\Lambda\bar{\Sigma}^{0}$ branching fractions within an SU(3) framework, considering variants with and without symmetry-breaking and  $\Lambda$-$\Sigma^0$ mixing effects. Due to the lack of experimental constraints, these predictions carry uncertainties comparable to their central values.  Experimental studies of  $\chi_{cJ}\to\Lambda\bar{\Sigma}^{0}$ decays would therefore provide crucial constraints to  improve the precision of theoretical predictions. Meanwhile, it can also deepen our understanding of the decay dynamics of the $\chi_{cJ}$ states~\cite{Ping:2004sh}, test theoretical approaches based on SU(3) flavor symmetry~\cite{PhysRevD.110.056007}, and probe possible $\Lambda$–$\Sigma^{0}$ mixing effects~\cite{Dalitz:1964es}.



In this work, using the world's largest $\psi(3686)$ dataset of 2.7 billion events collected with the BESIII detector~\cite{Ablikim_2024}, we present the first experimental searches for $\chi_{cJ}\to\Lambda\bar{\Sigma}^{0}$ decays.  Meanwhile, we investigate the analogous  $\eta_{c}\rightarrow \Lambda\bar{\Sigma}^{0}$ decay for the first time.
Charge-conjugate processes are implicitly included throughout this paper unless otherwise specified.

\section{BESIII DETECTOR AND MONTE CARLO SIMULATION}

The BESIII detector records symmetric $e^{+}e^{-}$ collisions provided by the BEPCII storage ring, which operates with a peak luminosity of $1.1 \times 10^{33}~\text{cm}^{-2}\text{s}^{-1}$ in the center-of-mass energy range from 1.84 to 4.95~GeV~\cite{2024138614}. 
A helium-based multilayer drift chamber (MDC), a plastic scintillator time-of-flight system (TOF), and a CsI (Tl) electromagnetic calorimeter (EMC) make up the cylindrical core of the BESIII detector, spanning 93\% of the full solid angle. They are mounted inside a superconducting solenoid magnet that generates a 1.0~T magnetic field. The solenoid is held in place by an octagonal magnetic flux return loop containing steel structures and resistive plate muon identification modules. The charged particle momentum resolution is 0.5\% at 1~GeV/c, and the electron identification in Bhabha scattering has a ${\rm d}E/{\rm d}x$ resolution of 6\%. The EMC measures photon energies with a resolution of 2.5\% (5\%) in the barrel (end-cap) region at 1~GeV.
The TOF barrel region has a time resolution of 68~ps, while the end-cap region achieves 110~ps. The end-cap TOF system was upgraded in 2015 with multi-gap resistive plate chamber technology, improving the time resolution to 60 ps. This enhancement benefits 83\% of the data used in this analysis~\cite{Li2017StudyOM,CAO2020163053,Guo2017TheSO}.
\par
Monte Carlo (MC) simulation is used to estimate the detection efficiency, optimize event selection criteria, and study potential backgrounds. The simulation is generated with a {\sc geant4}-based~\cite{AGOSTINELLI2003250} software package incorporating material properties, geometric descriptions, detector responses, and digitized models. 
A 2.7 billion event inclusive MC sample is used to investigate the background contributions. This sample includes the production of the $\psi(3686)$ resonance, the initial state radiation production of the $J/\psi$, and the continuum processes incorporated in {\sc kkmc}~\cite{JADACH2000260}. All particle decays are modeled with {\sc evtgen}~\cite{LANGE2001152} using branching fractions taken from the Particle Data Group (PDG)~\cite{10.1093/ptep/ptaa104}, while {\sc lundcharm}~\cite{PhysRevD.62.034003} is used for the unknown decay modes. Signal MC samples are generated to determine the detection efficiency and optimize selection criteria. The decays of $\psi(3686)\to\gamma\chi_{cJ}$ are modeled using the {\sc p2gcj} generator~\cite{Ping-Rong-Gang_2008}.
The decay of $\psi(3686)\to\gamma\eta_{c}$ is modeled using the {\sc jpe} model~\cite{Ping-Rong-Gang_2008}. The decays of $\chi_{cJ}\to\Lambda\bar{\Sigma}^{0}$, $\eta_{c}\to\Lambda\bar{\Sigma}^{0}$, $\Lambda\to p\pi^{-}$, and $\bar{\Sigma}^{0}\to\gamma\bar{\Lambda}$ are generated uniformly in phase space (PHSP).
Additionally, to estimate the expected number of background events and their shapes, the decays of $\chi_{cJ}\to\Sigma^{0}\bar{\Sigma}^{0}$, $\chi_{cJ}\to\Lambda\bar{\Lambda}$, and $\psi(3686)\to\gamma\Lambda\bar{\Sigma}^{0}$ are generated using the PHSP model, while  $\psi(3686)\to\Sigma^{0}\bar{\Sigma}^{0}$ is generated using the {\sc j2bb1} model~\cite{Ping-Rong-Gang_2008}.
\section{EVENT SELECTION AND DATA ANALYSIS}
The signal process $\psi(3686) \to \gamma_{1}\chi_{cJ}~(\eta_c) \to \gamma_{1}\gamma_{2} p\bar{p}\pi^+\pi^-$ requires the identification of four charged tracks with net charge zero and at least two photons.
Charged tracks, reconstructed from MDC hits, are selected using geometric criteria.  
The polar angle $\theta$ of each track relative to the beam axis must satisfy $|\cos\theta| < 0.93$, while the distance of closest approach to the interaction point (IP) is required to be within $|V_z| < 30~\text{cm}$ along the beam direction and $|V_{xy}| < 10~\text{cm}$ in the transverse plane.
\par
Proton-pion discrimination is achieved using momentum-dependent particle identification criteria optimized for each decay channel. For $\chi_{cJ} \to \Lambda\bar{\Sigma}^0$, charged tracks with laboratory frame momentum measured by the MDC $p > 0.60~\text{GeV}/c$ are identified as protons, and all others as pions. For $\eta_c \to \Lambda\bar{\Sigma}^0$, the proton identification threshold is lowered to $p > 0.38~\text{GeV}/c$.
\par
Photon candidates are selected based on energy deposition measured in the EMC, with $E_\gamma > 25~\text{MeV}$ in the barrel region ($|\cos\theta| < 0.80$) or $E_\gamma > 50~\text{MeV}$ in the end-cap region ($0.86 < |\cos\theta| < 0.92$).
Cluster timing is required to be within $[0, 700]~\text{ns}$ relative to the event start time to suppress electronic noise. 
To suppress showers associated with charged particles, we require the opening angle between the photon candidate and the nearest charged track to exceed $10^\circ$; this requirement is tightened to $20^\circ$ when the nearest track is an antiproton.
\par
\begin{figure}[htbp]
    \centering
    \includegraphics[width=0.5\linewidth]{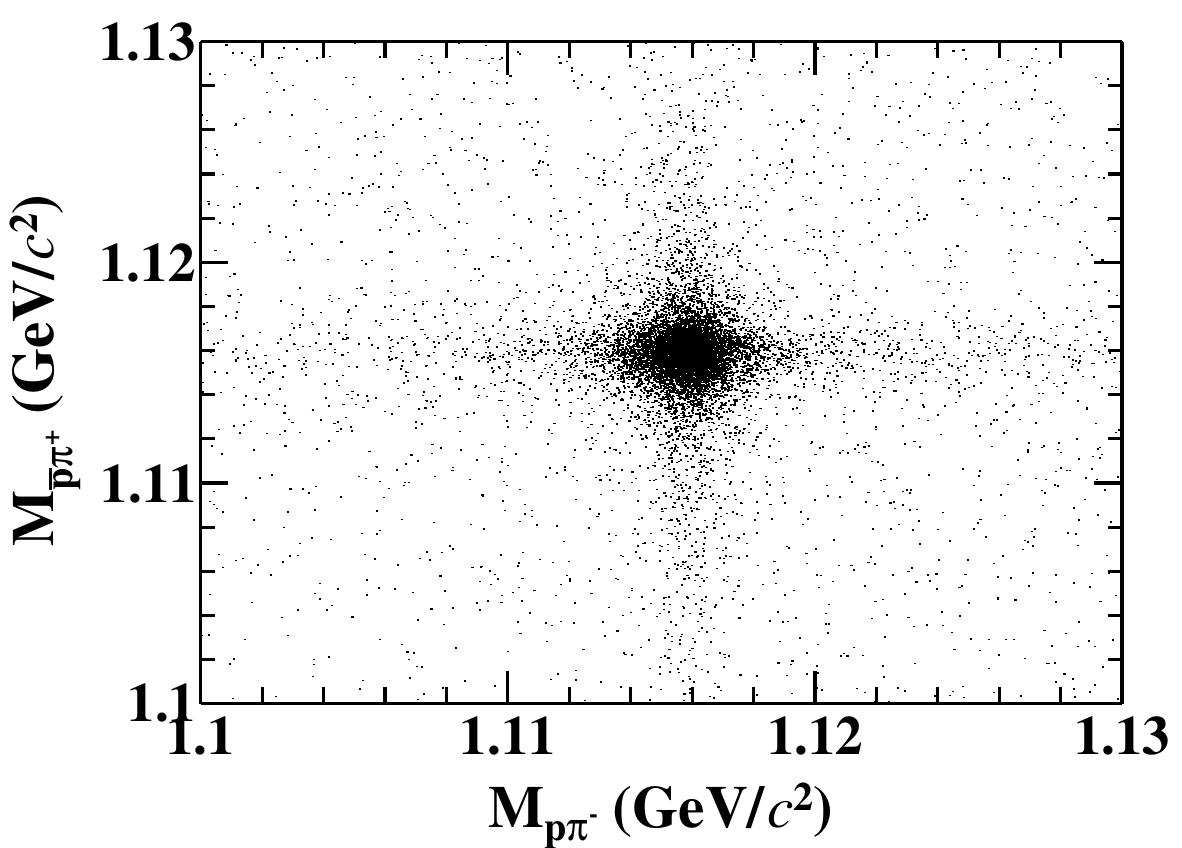}
    \put(-90,70){\textbf{(a)}}
    \includegraphics[width=0.5\linewidth]{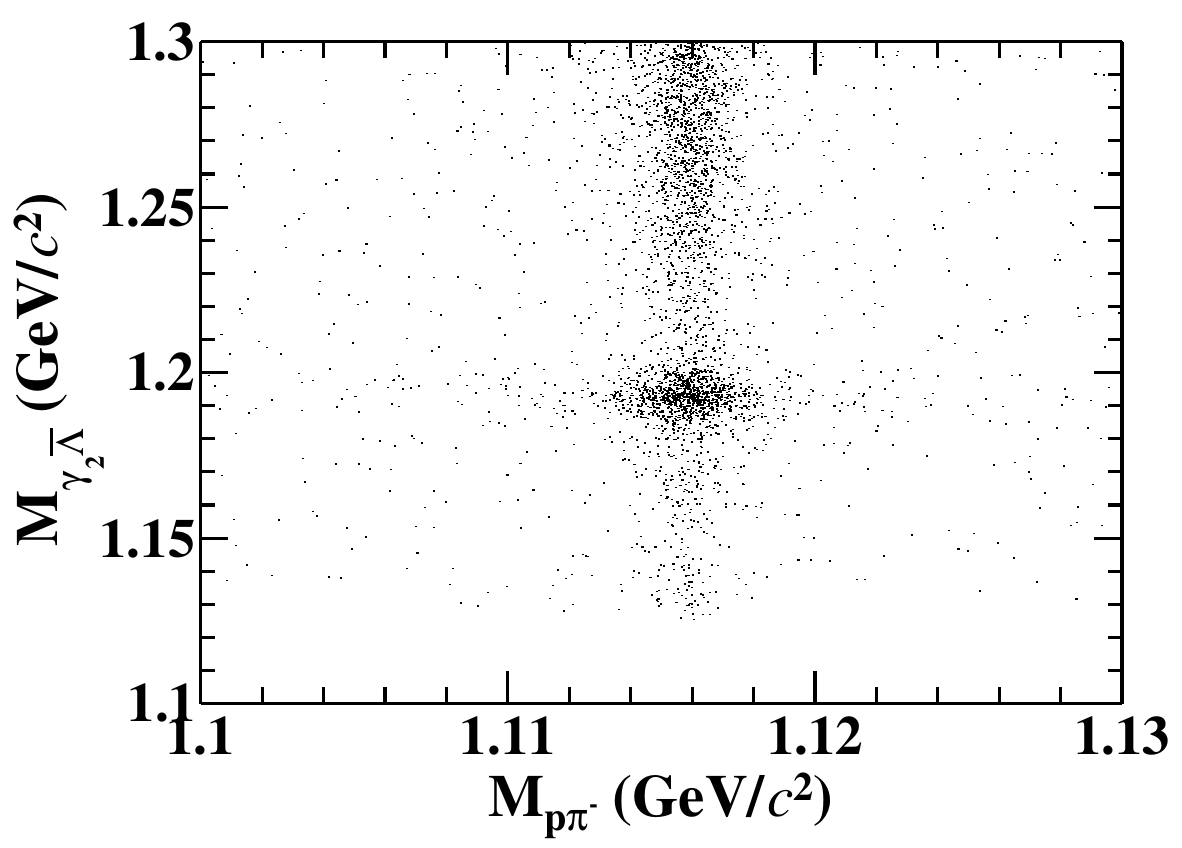}
    \put(-90,70){\textbf{(b)}}
    \caption{The distributions of M$_{p\pi^{-}}$ versus M$_{\bar{p}\pi^{+}}$ (a) and M$_{p\pi^{-}}$ versus M$_{\gamma_{2}\bar{\Lambda}}$ (b) for data.}
    \label{scatter_plots}
\end{figure}

The $\Lambda$ $(\bar{\Lambda})$ candidate is reconstructed from any $p\pi^{-}$ $(\bar{p}\pi^{+})$ combination that satisfies a successful secondary vertex fit~\cite{Min_2009}. To improve the momentum and energy resolution and reduce background contributions, a four-constraint (4C) kinematic fit imposing energy-momentum conservation is applied under the hypothesis of $\psi(3686)\to\gamma\gamma p\bar{p}\pi^{+}\pi^{-}$.  If there are more than two photons, the $\gamma\gamma$ combination with minimum $\chi_{\rm4C}^{2}$ is kept for further analysis.
The fit quality is required to satisfy $\chi^{2}_{\rm 4C} < 40$.
The distributions of M$_{p\pi^{-}}$ versus M$_{\bar{p}\pi^{+}}$ is shown in Fig.\ref{scatter_plots} (a), where clear $\Lambda$ pairs is observed.
To select the $\Lambda$ and $\bar{\Lambda}$ candidates, the mass window requirements of $\rm M_{p\pi^{-}/\bar{p}\pi^{+}} \in [1.110,\,1.120]~\mathrm{GeV}/c^2$ are applied.
The $\bar{\Sigma}^{0}$ candidates are reconstructed by combining each photon with the $\bar{\Lambda}$ candidate. 
The photon that minimizes $|\rm M_{\gamma\bar{\Lambda}} - M_{\bar{\Sigma}^0}^{\rm PDG}|$ is taken as $\gamma_2$ from $\bar{\Sigma}^0$ decay, where $\rm M_{\bar{\Sigma}^0}^{\rm PDG}$ is the nominal mass~\cite{10.1093/ptep/ptaa104}. While the other is assigned as $\gamma_1$ decaying from $\psi(3686)$ directly. 
The scatter plot of M$_{p\pi^{-}}$ versus M$_{\gamma_{2}\bar{\Lambda}}$ is shown in Fig.\ref{scatter_plots} (b), and clear accumulation around $\bar{\Sigma}^{0}$ is seen. 
An additional mass constraint of $\rm M_{\gamma_{2}\bar{\Lambda}} \in (1.178,\,1.208)~\mathrm{GeV}/c^2$ is applied to the chosen candidate.

\subsection{Search for $\bm{\chi_{cJ}\to\Lambda\bar{\Sigma}^{0}}$}
\begin{figure}[htbp]
    \includegraphics[width=0.5\linewidth]{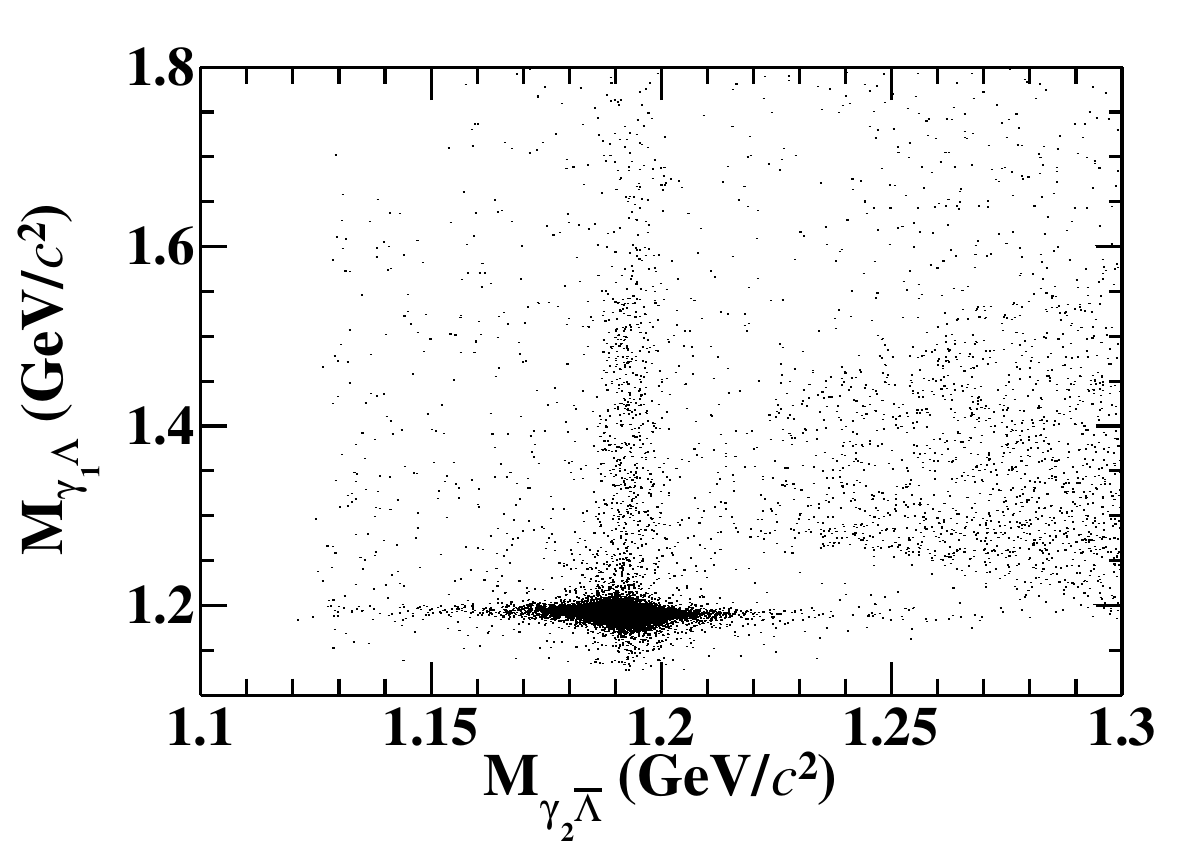}
    \put(-90,70){\textbf{(a)}}
    \includegraphics[width=0.5\linewidth]{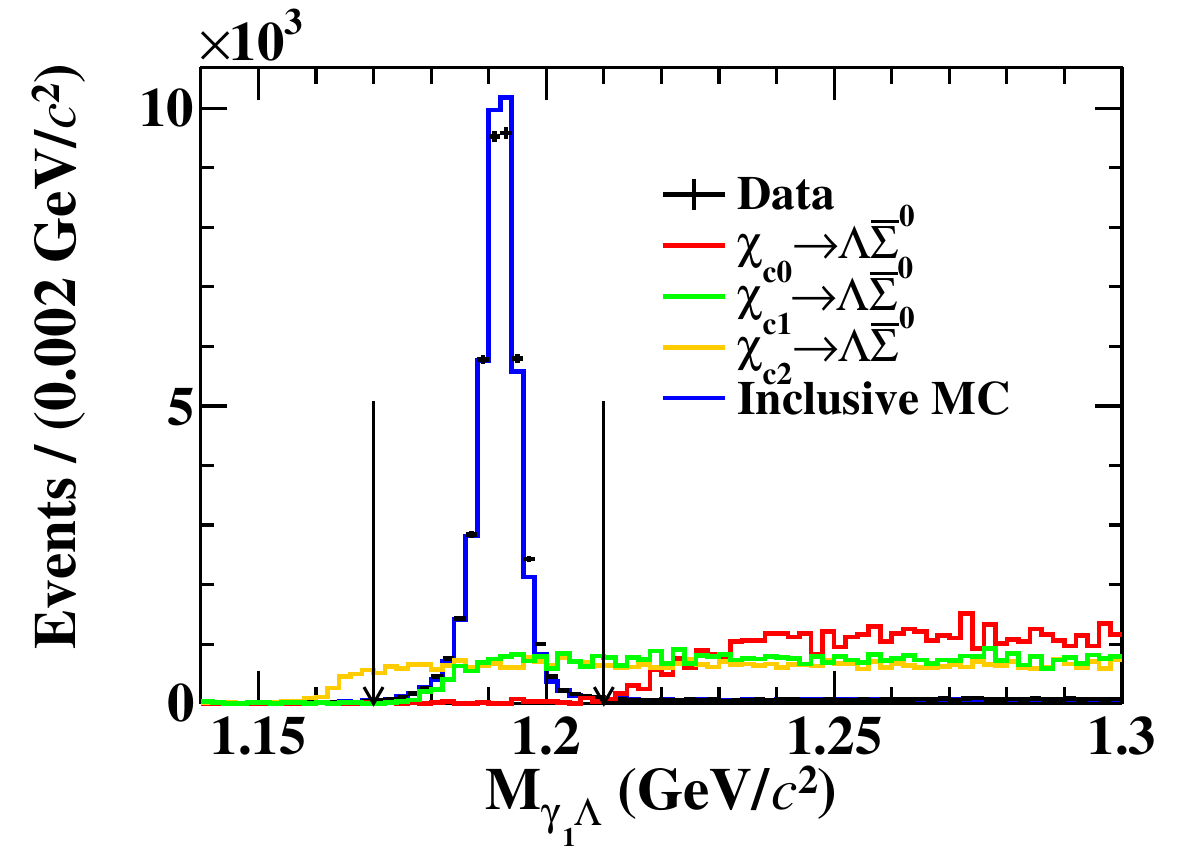}
    \put(-90,70){\textbf{(b)}}
    \caption{(a) The distribution of M$_{\gamma_{1}\Lambda}$ versus M$_{\gamma_{2}\bar{\Lambda}}$ for data. (b) The mass distribution of $\gamma_{1}\Lambda$. The red, green, and yellow curves are from MC simulations, the dots with error bars are data and the blue histogram is the inclusive MC sample.}
    \label{veto}
\end{figure}

Candidates for $\chi_{cJ}\to\Lambda\bar{\Sigma}^{0}$ are reconstructed from the combination of $\gamma_{2} p\bar{p}\pi^{+}\pi^{-}$. After the event selection, the same selection criteria are applied to an inclusive MC sample of 2.7 billion $\psi(3686)$ events to investigate the possible background contributions. 
The generic tool TopoAna~\cite{ZHOU2021107540} is used to study the remaining events.
The results indicate that the dominant backgrounds originate from $\psi(3686)\to\Sigma^{0}\bar{\Sigma}^{0}$, $\chi_{cJ}\to\Sigma^{0}\bar{\Sigma}^{0}$, $\chi_{cJ}\to\Lambda\bar{\Lambda}$, and $\psi(3686)\to\gamma\Lambda\bar{\Sigma}^{0}$.
The distributions of M$_{\gamma_{1}\Lambda}$ versus M$_{\gamma_{2}\bar{\Lambda}}$ and the distribution of $\rm M_{\gamma_{1}\Lambda}$ are shown in Fig.~\ref{veto}. An evident clustering of events near 1.19~GeV/$c^{2}$ is observed in this distribution, which corresponds to the background of $\psi(3686)\to\Sigma^{0}\bar{\Sigma}^{0}$. To suppress this background, the invariant mass of the $\gamma_{1}\Lambda$ system is required to lie outside the window (1.170, 1.210)~GeV/$c^{2}$.

In order to further reject the backgrounds of $\chi_{cJ}\to\Lambda\bar{\Lambda}$ and $\chi_{cJ}\to\Sigma^{0}\bar{\Sigma}^{0}$, the $\chi_{\rm 4C}^{2}$ for the $\gamma\gamma p\bar{p}\pi^{+}\pi^{-}$ ($\chi^{2}_{\gamma\gamma}$) hypothesis is required to be less than those for any of the $\gamma p\bar{p}\pi^{+}\pi^{-}$ ($\chi^{2}_{\gamma}$) and $\gamma\gamma\gamma p\bar{p}\pi^{+}\pi^{-}$ ($\chi^{2}_{\gamma\gamma\gamma}$) hypotheses: $\chi_{\gamma\gamma}^{2}<\chi_{\gamma\gamma\gamma}^{2}$ and $\chi_{\gamma\gamma}^{2}<\chi_{\gamma}^{2}$. 
\par
\begin{figure*}
    \centering
    \includegraphics[width=1.0\linewidth]{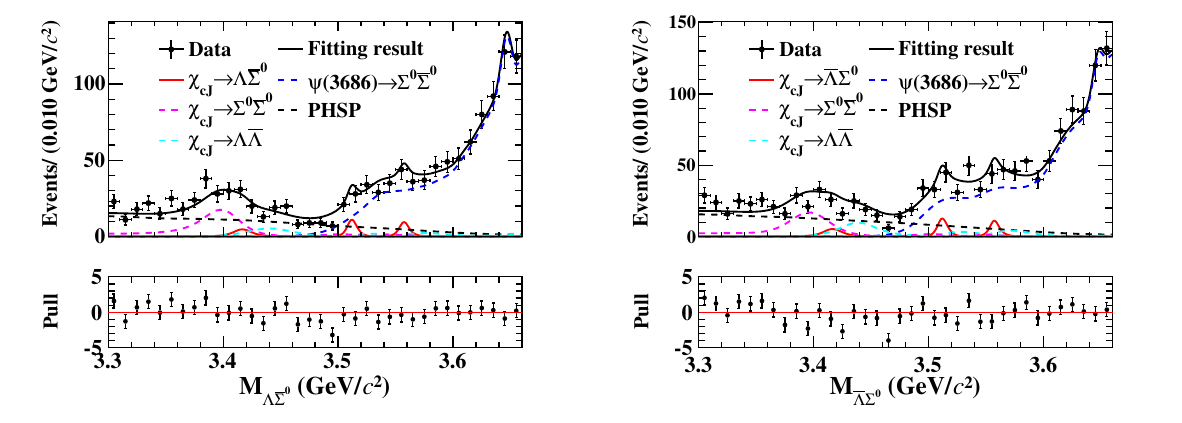}
    \put(-315,160){\textbf{(a)}}
    \put(-60,160){\textbf{(b)}}
    \caption{Simultaneous fit to the mass distributions of $\Lambda\bar{\Sigma}^{0}$ and $\bar{\Lambda}\Sigma^{0}$ in the $\chi_{cJ}$ mass region. The dots with error bars are data, the pink dotted lines are the background of $\chi_{cJ}\to\Sigma^{0}\bar{\Sigma}^{0}$, the cyan dotted lines indicate the background of $\chi_{cJ}\to\Lambda\bar{\Lambda}$, the blue dotted lines are the background of $\psi(3686)\to\Sigma^{0}\bar{\Sigma}^{0}$, the black dotted lines are the background of $\psi(3686)\to\gamma\Lambda\bar{\Sigma}^{0}$, and the red lines indicate the signal shape.}
    \label{fit_chi_cj}
\end{figure*}

After the above selection criteria, the mass distributions of $\Lambda\bar{\Sigma}^{0}$ are shown in Fig.~\ref{fit_chi_cj} in the region of interest (3.30, 3.66)~GeV/$c^{2}$. 
The mass distribution can be almost described with backgrounds from $\psi(3686)\to\Sigma^{0}\bar{\Sigma}^{0}$, $\chi_{cJ}\to\Sigma^{0}\bar{\Sigma}^{0}$, $\chi_{cJ}\to\Lambda\bar{\Lambda}$, and $\psi(3686)\to\gamma\Lambda\bar{\Sigma}^{0}$.
An unbinned maximum likelihood fit is performed to the $\Lambda\bar{\Sigma}^{0}$ and $\bar{\Lambda}\Sigma^{0}$ invariant mass distributions simultaneously. The branching fraction $\mathcal{B}(\chi_{cJ}\to\Lambda\bar{\Sigma}^{0})$ is calculated as:
\begin{align}
 \mathcal{B}(\chi_{cJ}\to\Lambda\bar{\Sigma}^{0})= A(\frac{N_{\rm sig}^{\chi_{cJ}\to\Lambda\bar{\Sigma}^{0}}}{\epsilon^{\chi_{cJ}\to\Lambda\bar{\Sigma}^{0}}}+\frac{N_{\rm sig}^{\chi_{cJ}\to\bar{\Lambda}\Sigma^{0}}}{\epsilon^{\chi_{cJ}\to\bar{\Lambda}\Sigma^{0}}}),
 \label{formula_chi_cj}
\end{align}
where $A$ is defined as $1/(N_{\rm tot}\times\mathcal{B}_{\rm inter})$.
$N_{\rm tot}$ is the total number of $\psi(3686)$ events in data, and
$\mathcal{B}_{\rm inter}$ is the product of the branching fractions
$\mathcal{B}(\psi(3686)\to\gamma\chi_{cJ})\times
\mathcal{B}(\bar{\Sigma}^{0}\to\gamma\bar{\Lambda})\times
\mathcal{B}(\Lambda\to p\pi^{-})^{2}$.
$N_{\rm sig}^{\chi_{cJ}\to\Lambda\bar{\Sigma}^{0}}$ and
$N_{\rm sig}^{\chi_{cJ}\to\bar{\Lambda}\Sigma^{0}}$ are the signal yields of the two charge-conjugate decay modes, while $\epsilon^{\chi_{cJ}\to\Lambda\bar{\Sigma}^{0}}$ and
$\epsilon^{\chi_{cJ}\to\bar{\Lambda}\Sigma^{0}}$ are the corresponding detection efficiencies, which are determined with the dedicated MC samples.
The $\chi_{cJ}$ signal, the background from $\psi(3686)\to\Sigma^{0}\bar{\Sigma}^{0}$, $\chi_{cJ}\to\Sigma^{0}\bar{\Sigma}^{0}$, $\chi_{cJ}\to\Lambda\bar{\Lambda}$, and the PHSP shape from $\psi(3686)\to\gamma\Lambda\bar{\Sigma}^{0}$ are described with the MC-simulated shapes. 
The contributions of $\psi(3686)\to\Sigma^{0}\bar{\Sigma}^{0}$, $\chi_{cJ}\to\Sigma^{0}\bar{\Sigma}^{0}$, and $\chi_{cJ}\to\Lambda\bar{\Lambda}$ are fixed to their yields which are calculated with the total number of $\psi(3686)$ events in data~\cite{Ablikim_2024}, the branching fractions taken from the PDG~\cite{10.1093/ptep/ptaa104}, and the reconstruction efficiencies derived from dedicated MC simulations of background processes.
The statistical significance is calculated from the change in the negative log-likelihood function, $-2\Delta\ln\mathcal{L}$, between the fits with and without the signal component, accounting for the difference in the number of degrees of freedom. 
The resulting significances are $0\sigma$ for $\chi_{c0}$, $1.2\sigma$ for $\chi_{c1}$, and $0.3\sigma$ for $\chi_{c2}$.




A Bayesian approach~\cite{Rover:2011zq} is employed to determine the ULs of $\mathcal{B}(\chi_{cJ}\to\Lambda\bar{\Sigma}^{0})$ at the 90\% confidence level (CL). The normalized likelihood defined as $\mathcal{L}(N)=exp(-[\mathcal{S}(N)-\mathcal{S}_{\rm min}])$, is obtained from the fits, where $\mathcal{S}_{\rm min}$ is the minimum value of the negative log-likelihood~\cite{BESIII:2024ddb}. A series of fits are performed with the branching fraction $\mathcal{B}(\chi_{cJ}\to\Lambda\bar{\Sigma}^{0})$ fixed at values from $4\times10^{-8}$ to $8\times10^{-6}$ in steps of $4\times10^{-8}$. For each fixed branching fraction, the corresponding likelihood value is obtained.
The ULs on the branching fractions at the 90\% CL are then determined by integrating the likelihood distribution from zero to 90\% of the total area in the physical region ($\mathcal{B}\geq 0$).
Additionally, to incorporate the multiplicative systematic uncertainties into the UL calculation, the likelihood distribution is smeared by a Gaussian function with a mean of zero and a width equal to the total multiplicative uncertainty as in Eq.~\eqref{result}~\cite{stenson2006exactsolutionincorporatingmultiplicative},
\begin{equation}
    L^{\prime}(\mathcal{B})=\int_0^1 L\left(\frac{S}{\hat{S}} \mathcal{B}\right) \exp \left[-\frac{(S-\hat{S})^{2}}{2 \sigma_{S}^2}\right] d S,
\label{result}
\end{equation}
where $L^{\prime}$ is the smeared likelihood distribution, $L$ represents the original likelihood distribution without systematic uncertainties, $S$ is the detection efficiency, $\hat{S}$ denotes the nominal detection efficiency averaged over the charge-conjugate processes, and $\sigma_{S}$ is the multiplicative systematic uncertainty (see Section~\ref{systematic} below for details).

\subsection{Search for $\bm{\eta_{c}\to\Lambda\bar{\Sigma}^{0}}$}
\begin{figure*}
    \centering
    \includegraphics[width=1.0\linewidth]{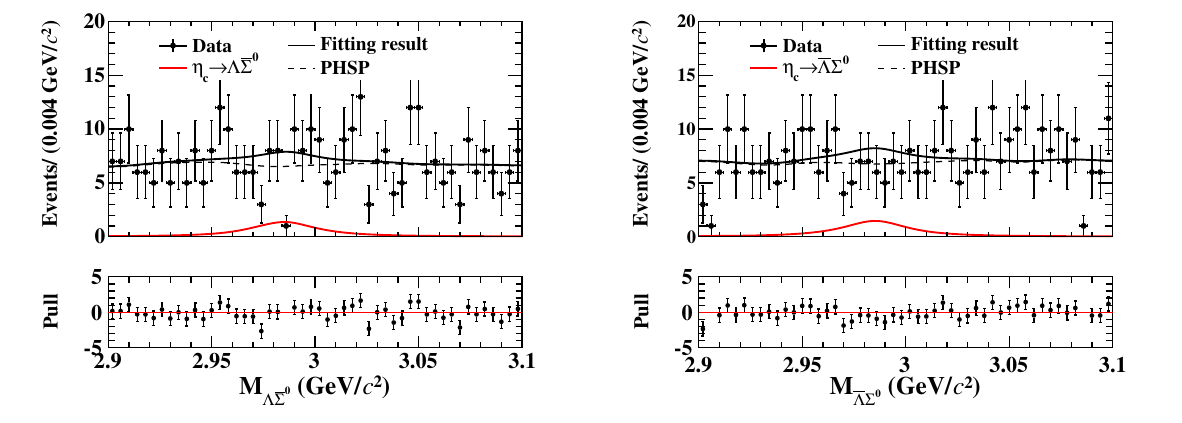}
    \put(-315,160){\textbf{(a)}}
    \put(-60,160){\textbf{(b)}}
    \caption{Simultaneous fit for the mass distributions of $\Lambda\bar{\Sigma}^{0}$ and $\bar{\Lambda}\Sigma^{0}$ in the $\eta_c$ mass region. The dots with error bars are data, the black lines are the fitting result, the black dotted lines are the background of $\psi(3686)\to\gamma\Lambda\bar{\Sigma}^{0}$, and the red lines indicate the signal shape.}
    \label{fit_eta_c}
\end{figure*}

Similar to the $\chi_{cJ}$ selection, backgrounds from $\eta_{c} \to \Lambda\bar{\Lambda}$ and $\eta_{c} \to \Sigma^{0}\bar{\Sigma}^{0}$ are suppressed by requiring the $\chi^{2}$ from the 4C kinematic fit under the $\gamma\gamma p\bar{p}\pi^{+}\pi^{-}$ hypothesis to be less than that under the $\gamma p\bar{p}\pi^{+}\pi^{-}$ and $\gamma\gamma\gamma p\bar{p}\pi^{+}\pi^{-}$ hypotheses: i.e., $\chi^{2}_{\gamma\gamma} < \chi^{2}_{\gamma}$ and $\chi^{2}_{\gamma\gamma} < \chi^{2}_{\gamma\gamma\gamma}$.
After applying these selection criteria, no evident $\eta_{c}$ signal is observed in the mass distributions of $\Lambda\bar{\Sigma}^{0}$ within the region (2.90, 3.10)~GeV/$c^{2}$, as shown in Fig.~\ref{fit_eta_c}. An unbinned maximum likelihood fit is performed on the $\Lambda\bar{\Sigma}^{0}$ and $\bar{\Lambda}\Sigma^{0}$ invariant mass distribution to set the UL on the branching fraction at the 90\% CL simultaneously. Similar to the search for $\chi_{cJ}\to\Lambda\bar{\Sigma}^{0}$, the branching
fraction $\mathcal{B}(\eta_{c}\to\Lambda\bar{\Sigma}^{0})$ is calculated as:
\begin{equation}
 \mathcal{B}(\eta_{c}\to\Lambda\bar{\Sigma}^{0})=A^{\prime}(\frac{N_{\rm sig}^{\eta_{c}\to\Lambda\bar{\Sigma}^{0}}}{\epsilon^{\eta_{c}\to\Lambda\bar{\Sigma}^{0}}}+\frac{N_{\rm sig}^{\eta_{c}\to\bar{\Lambda}\Sigma^{0}}}{\epsilon^{\eta_{c}\to\bar{\Lambda}\Sigma^{0}}}).
\label{formula_eta_c}
\end{equation}
Here, $A^{\prime} = 1/(N_{\rm tot}\times\mathcal{B}_{\rm inter}^{\prime})$ and the $\mathcal{B}_{\rm inter}^{\prime}$ represents the product branching fraction $\mathcal{B}(\psi(3686)\to\gamma\eta_{c})\times
\mathcal{B}(\bar{\Sigma}^{0}\to\gamma\bar{\Lambda})\times
\mathcal{B}(\Lambda\to p\pi^{-})^{2}$. $N_{\rm sig}^{\eta_{c}\to\Lambda\bar{\Sigma}^{0}}$ and
$N_{\rm sig}^{\eta_{c}\to\bar{\Lambda}\Sigma^{0}}$ denote the signal yields of the two charge-conjugate decay modes, and $\epsilon^{\eta_{c}\to\Lambda\bar{\Sigma}^{0}}$ and
$\epsilon^{\eta_{c}\to\bar{\Lambda}\Sigma^{0}}$ denote the corresponding detection efficiencies obtained from the MC simulation.
The possible $\eta_{c}$ signal is described with MC-simulated shape. The smooth background is described with a PHSP shape for $\psi(3686)\to\gamma\Lambda\bar{\Sigma}^{0}$, obtained from the dedicated MC simulations. 
Following the same procedure as for the  $\chi_{cJ}\to\Lambda\bar{\Sigma}^{0}$ analysis, a Bayesian approach is applied to set the UL on the branching fraction at the 90\% CL. The branching fraction $\mathcal{B}(\eta_{c}\to\Lambda\bar{\Sigma}^{0})$ is fixed at values from $1\times10^{-6}$ to $2\times10^{-4}$ in steps of $1\times10^{-6}$.
The UL on the branching fraction is determined by integrating the likelihood curve from zero to 90\% of the area in the physical region. The multiplicative systematic uncertainty is also considered with Eq.~\ref{result}.

\section{SYSTEMATIC UNCERTAINTY}
\label{systematic}
Systematic uncertainties are classified into two categories: additive uncertainties, which affect the observed event yield, and multiplicative uncertainties, which affect the detection efficiencies. For a pair of charge-conjugate processes, the larger of the two multiplicative uncertainties from a given source is assigned as the systematic uncertainty for the combined process.
The multiplicative uncertainties are summarized in Table~\ref{total_sys}. 
\begin{table}
    \centering
    \caption{Summary of multiplicative systematic uncertainties (in \%) for the upper limit measurements.}

    \begin{tabular}{@{}l|cccc@{}}
         \hline\hline
         Source&$\chi_{c0}$&$\chi_{c1}$&$\chi_{c2}$&$\eta_{c}$\\
         \hline
        \makecell[l]{Tracking, photon detection \\ and $\Lambda(\bar{\Lambda})$ construction} & 2.3 & 2.3 & 2.3 & 2.3 \\
         Kinematic fit& 1.2&1.4&1.3&1.5\\
         Quoted branching fractions &2.8&3.2&2.9&14.0\\
         $N_{\rm tot}$&0.5&0.5&0.5&0.5\\
         MC model&1.6&2.5&2.7&1.1\\
         $\Lambda,\bar{\Lambda}$ mass window &0.6&0.6&0.6&0.6\\
         $\bar{\Sigma}^{0}$ mass window&0.1&0.1&0.1&0.1\\
         Veto $\psi(3686)\to\Sigma^{0}\bar{\Sigma}^{0}$&0.1&0.1&0.1&-\\
         \hline
         Total&4.2&4.9&4.8&14.3\\
         \hline\hline
    \end{tabular}
    \label{total_sys}
\end{table}
\begin{figure}[htbp]
    \centering
    \includegraphics[width=0.49\linewidth]{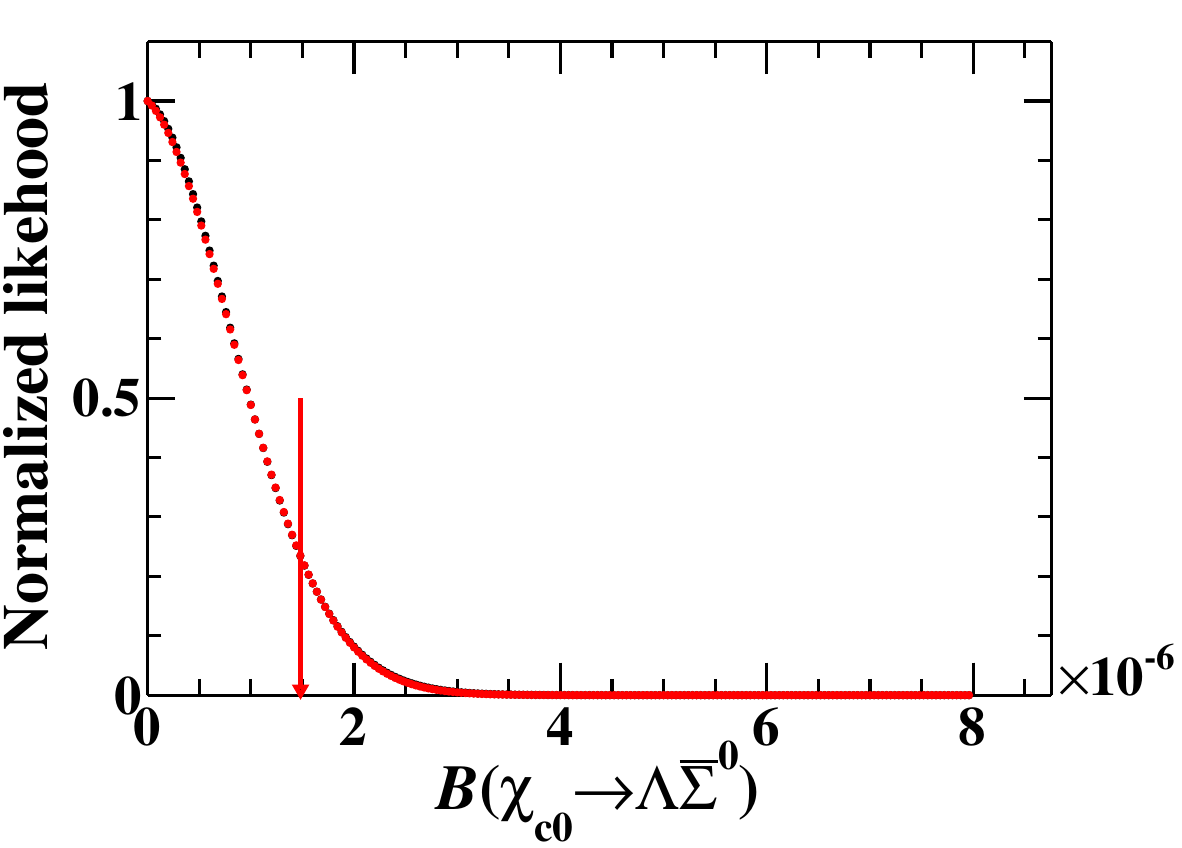}
    \includegraphics[width=0.49\linewidth]{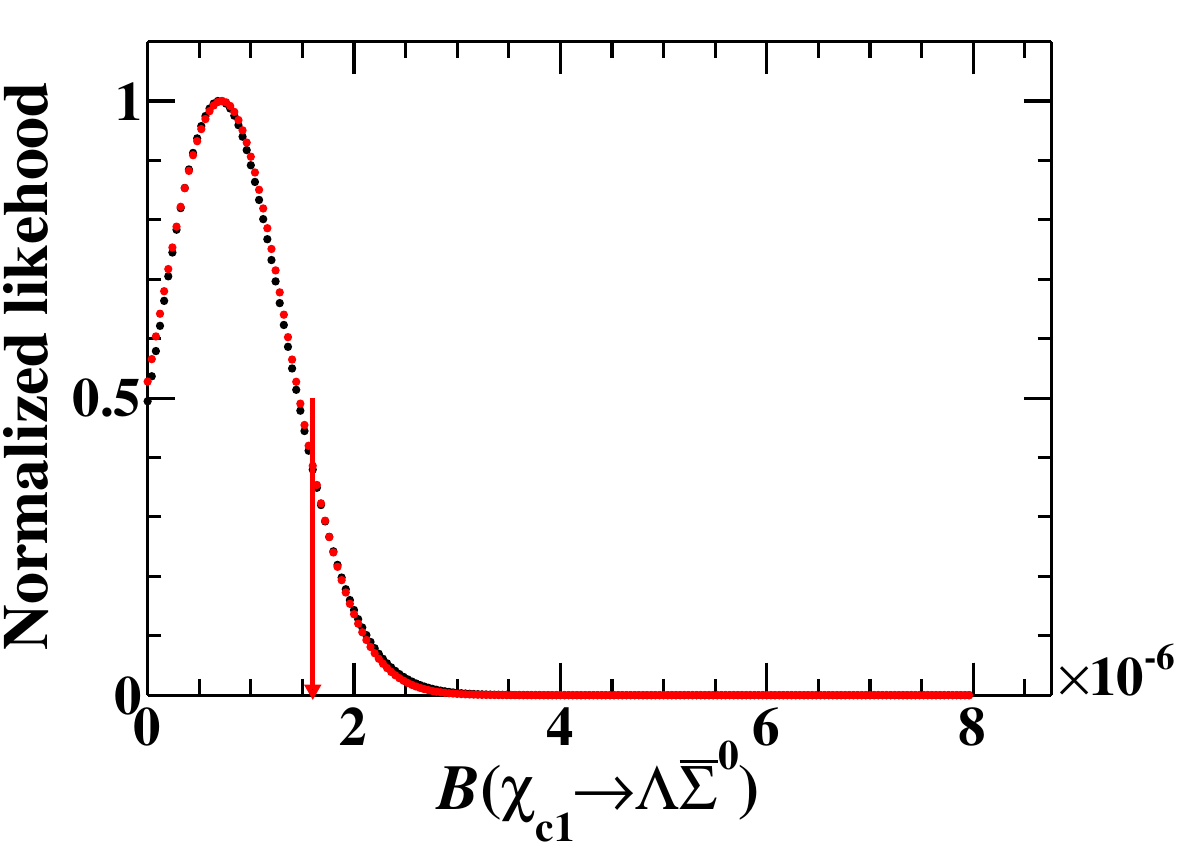}
    \includegraphics[width=0.49\linewidth]{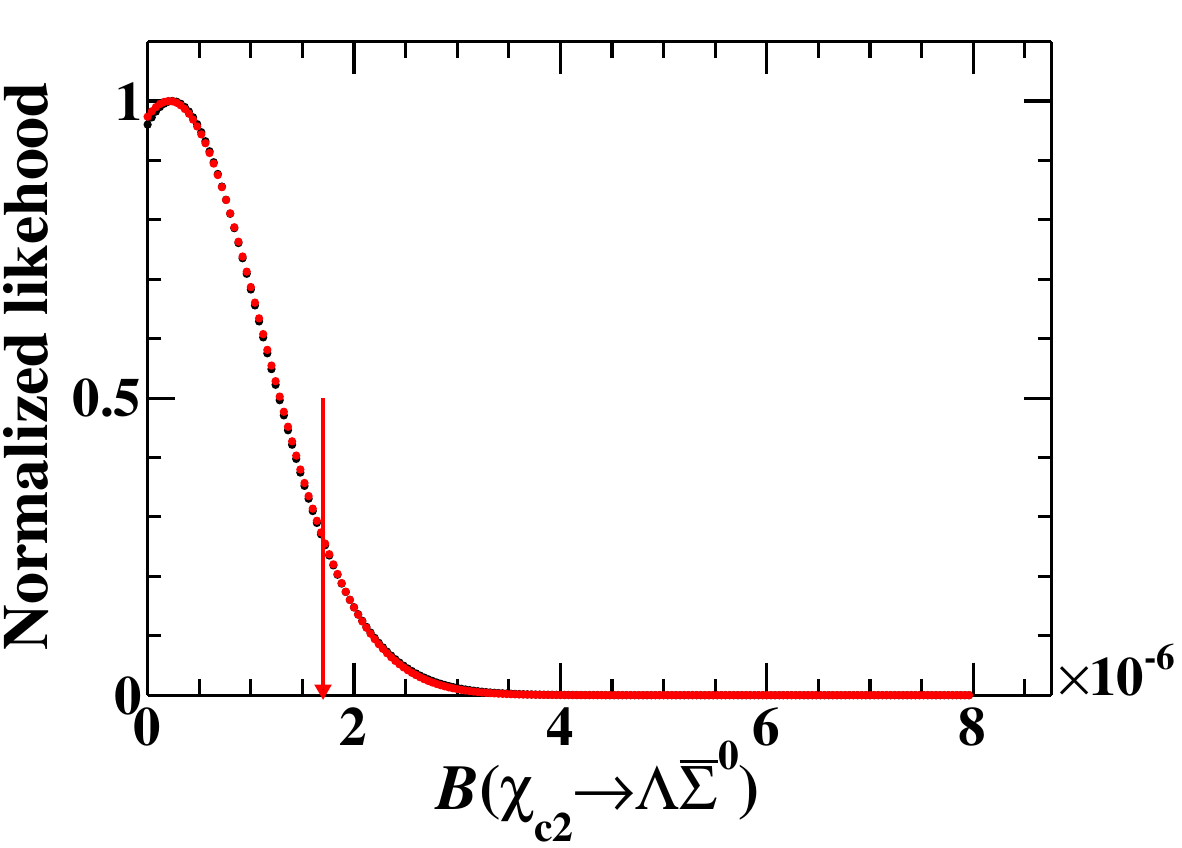}
    \includegraphics[width=0.49\linewidth]{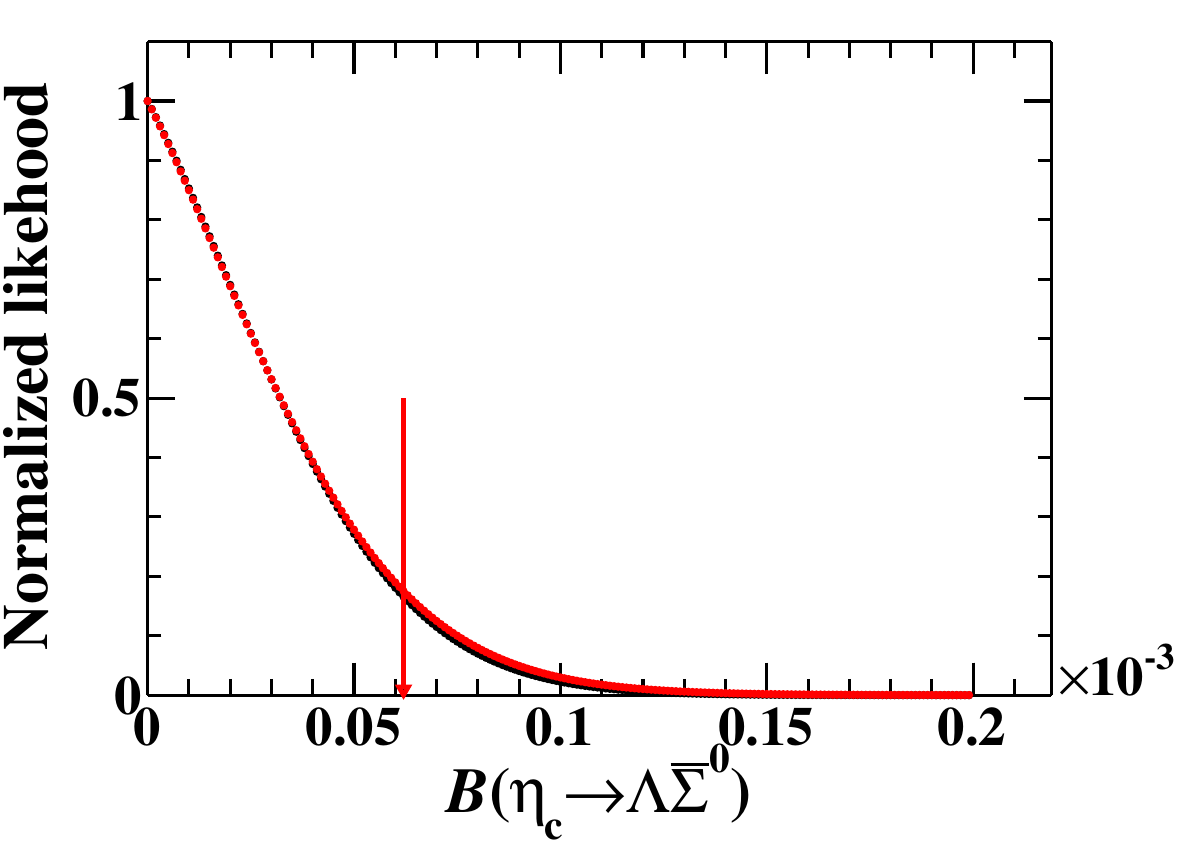}
    \caption{Normalized likelihood distributions before (black dots) and after (red dots) considering the multiplication uncertainties for $\chi_{cJ}\to\Lambda\bar{\Sigma}^{0}$ and $\eta_{c}\to\Lambda\bar{\Sigma}^{0}$. The black dotted-dashed lines are the initial curves, the red dotted-dashed lines are the results after convolving with a Gaussian function for multiplication uncertainties, and the red arrows indicate the ULs on the branching fractions at the 90\% CL.}
    \label{smear}
\end{figure}
\par
\begin{table*}[!htbp]
\caption{
ULs on the branching fractions of $\chi_{cJ}\to\Lambda\bar{\Sigma}^{0}$ from this work compared with theoretical predictions.  Model I: the SU(3) flavor symmetry and breaking contributions are considered. Model II: the SU(3) flavor symmetry and breaking, as well as the $\Sigma^{0}-\Lambda$ mixing are all included, where the mixing angle is set as $\alpha=0.015\pm0.001$. Model I$^\prime$ (Model II$^\prime$) is defined similar to Model I (Model II) where the SU(3) breaking contribution is limited at 20\% of the leading SU(3) symmetric amplitude. The uncertainties are combined from both statistical and systematic effects.} 
\begin{tabular}{@{}lcccc@{}}
    \hline\hline
    Model&$\chi_{c0}\to\Lambda\bar{\Sigma}^{0}$&$\chi_{c1}\to\Lambda\bar{\Sigma}^{0}$&$\chi_{c2}\to\Lambda\bar{\Sigma}^{0}$\\
    \hline
    \uppercase\expandafter{\romannumeral1}~\cite{PhysRevD.110.056007} &$(0.37\pm0.37)\times10^{-4}$&$(6.34\pm6.34)\times10^{-5}$ &$(3.65\pm3.65)\times10^{-5}$\\
    \uppercase\expandafter{\romannumeral2}~\cite{PhysRevD.110.056007} &$(0.80\pm0.80)\times10^{-4}$&$(11.84\pm11.83)\times10^{-5}$ & $(7.25\pm7.25)\times10^{-5}$\\
    \uppercase\expandafter{\romannumeral1}$^\prime$~\cite{PhysRevD.110.056007} &$(0.13\pm0.13)\times10^{-4}$&$(0.06\pm0.06)\times10^{-5}$&$(0.12\pm0.12)\times10^{-5}$\\
    \uppercase\expandafter{\romannumeral2}$^\prime$~\cite{PhysRevD.110.056007} &$(0.26\pm0.26)\times10^{-4}$&$(0.18\pm0.18)\times10^{-5}$&$(0.27\pm0.27)\times10^{-5}$\\
    \hline
    This work&$<1.5\times 10^{-6}$&$<1.6\times 10^{-6}$&$<1.7\times 10^{-6}$\\
    \hline\hline
    \label{summary}
\end{tabular}
\end{table*}
For both $\chi_{cJ}\to\Lambda\bar{\Sigma}^{0}$ and $\eta_{c}\to\Lambda\bar{\Sigma}^{0}$ decays, the multiplicative systematic uncertainties mainly come from: tracking, photon detection, $\Lambda$ ($\bar{\Lambda}$) reconstruction, quoted branching fractions, the total number of $\psi(3686)$ events, kinematic fit, MC model, and mass window. An additional contribution from the veto of $\psi(3686)\to\Sigma^{0}\bar{\Sigma}^{0}$ background is considered for the $\chi_{cJ}$ decays only.
The systematic uncertainties from tracking, photon detection and $\Lambda$ ($\bar{\Lambda}$) reconstruction are estimated together with a control sample of $\psi(3686)\to\Sigma^{0}\bar{\Sigma}^{0}$ due to the similarity in kinematics and event selections, and quoted as 2.3\% from Ref.~\cite{PhysRevD.95.052003}. 
The systematic uncertainties on the quoted branching fractions are taken from the PDG~\cite{10.1093/ptep/ptaa104} values, which are 2.8\%, 3.2\%, 2.9\%, and 14.0\% for $\chi_{c0}$, $\chi_{c1}$, $\chi_{c2}$, and $\eta_{c}$, respectively.
The total number of $\psi(3686)$ events is $(2712.4\pm14.3)\times10^{6}$~\cite{Ablikim_2024}, corresponding to a relative uncertainty of 0.5\%. The uncertainty associated with the kinematic fit is estimated by comparing the efficiencies with and without the helix parameter correction~\cite{PhysRevD.87.012002}. The resulting uncertainties are 1.2\%, 1.4\%, 1.3\%, and 1.5\% for $\chi_{c0}$, $\chi_{c1}$, $\chi_{c2}$, and $\eta_{c}$, respectively. To evaluate the uncertainty from the MC model, an alternative MC sample with {\sc chijbb1} and {\sc pbb1}~\cite{Ping-Rong-Gang_2008} models is generated. The difference in efficiencies between the nominal and alternative MC models is taken as the uncertainty.
A control sample of  $\psi(3686)\to\Sigma^{0}\bar{\Sigma}^{0}$ is used to evaluate the systematic uncertainties associated with the mass windows and the corresponding veto requirements. The uncertainties are estimated by varying the relevant mass windows by $\pm$ 2~$\text{MeV}/c^2$, and the resulting difference in the measured branching fraction is assigned as the corresponding uncertainty.
All multiplicative systematic uncertainties are added in quadrature to obtain the total multiplicative uncertainty.
\par
The additive systematic uncertainties for both $\chi_{cJ}\to\Lambda\bar{\Sigma}^{0}$ and $\eta_{c}\to\Lambda\bar{\Sigma}^{0}$ decays primarily arise from three sources: (1) the signal parameterization, (2) the background modeling, and (3) the fit range. 
For both processes, the systematic uncertainty from the signal parameterization is estimated by replacing the nominal MC shape with a double-Gaussian function, whose parameters are determined by fitting the signal MC sample.
The systematic uncertainties due to the background modeling are assessed differently: for $\chi_{cJ}$ by varying the number of events of the background processes $\psi(3686)\to\Sigma^{0}\bar{\Sigma}^{0}$, $\chi_{cJ}\to\Sigma^{0}\bar{\Sigma}^{0}$, and $\chi_{cJ}\to\Lambda\bar{\Lambda}$ by $\pm$ 1 standard deviation; while for $\eta_{c}$, they are evaluated by replacing the PHSP background model with a first-order polynomial function. The systematic uncertainty associated with the fit range is evaluated by varying the fitted mass interval by $\pm$ 10~MeV/$c^{2}$. The most conservative UL result is considered as the final result.

After accounting for the additive uncertainties, Eq.~\ref{result} is employed to incorporate the multiplicative uncertainties. Figure~\ref{smear} illustrates the normalized likelihood distributions and the ULs of the branching fractions are determined at the 90\% CL.
\par

\section{SUMMARY}
Based on a sample of $(2712.4 \pm 14.3)\times10^6$ $\psi(3686)$ events collected with the BESIII detector, the ULs on the $\mathcal{B}(\chi_{cJ}\to\Lambda\bar{\Sigma}^{0})$ and $\mathcal{B}(\eta_{c}\to\Lambda\bar{\Sigma}^{0})$ at the 90\% CL are set to be $1.5\times10^{-6}$, $1.6\times10^{-6}$, $1.7\times10^{-6}$, and $6.2\times10^{-5}$ for $\chi_{c0}$, $\chi_{c1}$, $\chi_{c2}$, and $\eta_{c}$, respectively.
These results are consistent with expectations based on the helicity selection rule for $\chi_{c0}$ decays.
Table~\ref{summary} compares the ULs obtained in this work with the theoretical predictions for the $\chi_{cJ}$ branching fractions from Ref.~\cite{PhysRevD.110.056007}. 
Notably, most of the predicted central values are significantly larger than our measurements.
This comparison highlights the need for improved theoretical constraints on non-perturbative quantum chromodynamics parameters governing isospin-violating charmonium decays.
\section{ACKNOWLEDGMENTS}

\textbf{Acknowledgement}

The BESIII Collaboration thanks the staff of BEPCII (https://cstr.cn/31109.02.BEPC) and the IHEP computing center for their strong support. This work is supported in part by National Key R\&D Program of China under Contracts Nos. 2025YFA1613900, 2023YFA1606000, 2023YFA1606704; Beijing Natural Science Foundation (BJNSF) under Contract No. JQ22002; National Natural Science Foundation of China (NSFC) under Contracts Nos. 11635010, 11935015, 11935016, 11935018, 12025502, 12035009, 12035013, 12061131003, 12105100, 12192260, 12192261, 12192262, 12192263, 12192264, 12192265, 12221005, 12225509, 12235017, 12342502, 12361141819; the Chinese Academy of Sciences (CAS) Large-Scale Scientific Facility Program; the Strategic Priority Research Program of Chinese Academy of Sciences under Contract No. XDA0480600; CAS under Contract No. YSBR-101; 100 Talents Program of CAS; The Institute of Nuclear and Particle Physics (INPAC) and Shanghai Key Laboratory for Particle Physics and Cosmology; ERC under Contract No. 758462; German Research Foundation DFG under Contract No. FOR5327; Istituto Nazionale di Fisica Nucleare, Italy; Knut and Alice Wallenberg Foundation under Contracts Nos. 2021.0174, 2021.0299, 2023.0315; Ministry of Development of Turkey under Contract No. DPT2006K-120470; National Research Foundation of Korea under Contract No. NRF-2022R1A2C1092335; National Science and Technology fund of Mongolia; Polish National Science Centre under Contract No. 2024/53/B/ST2/00975; STFC (United Kingdom); Swedish Research Council under Contract No. 2019.04595; U. S. Department of Energy under Contract No. DE-FG02-05ER41374;


\bibliographystyle{apsrev4-2}
\bibliography{reference}

\end{document}